\documentclass[aip, reprint, amsmath, amssymb, superscriptaddress,floatfix,twocolumn]{revtex4-1}
\usepackage{amsmath}
\usepackage{float}

\usepackage{graphicx}
\usepackage{dcolumn}
\usepackage{bm}
\usepackage{bbm}
\usepackage{sidecap}
\usepackage{braket}
\usepackage{color}
\usepackage{amssymb}
\usepackage{MnSymbol}
\usepackage{csquotes}

\newcommand{\Fig}[1]{Fig.\,\ref{#1}}
\newcommand{\Eq}[1]{Eq.\,(\ref{#1})}

\newcommand{\be}{\begin{equation}}
\newcommand{\ee}{\end{equation}}
\newcommand{\bea}{\begin{eqnarray}}
\newcommand{\eea}{\end{eqnarray}}

\newcommand{\RNum}[1]{\uppercase\expandafter{\romannumeral #1\relax}}

\begin{document}
\title{Hierarchical equations of motion approach to hybrid fermionic and
bosonic environments: Matrix product state formulation in twin space}

\author{Yaling Ke}
\affiliation{
	Institute of Physics, Albert-Ludwig University Freiburg, Hermann-Herder-Strasse 3, 79104 Freiburg, Germany
}

\author{Raffaele Borrelli}
\affiliation{
DISAFA, Università di Torino, I-10095 Grugliasco, Italy}

\author{Michael Thoss}
\affiliation{
Institute of Physics, Albert-Ludwig University Freiburg, Hermann-Herder-Strasse 3, 79104 Freiburg, Germany
}
\affiliation{
EUCOR Centre for Quantum Science and Quantum Computing, Albert-Ludwig
University Freiburg, Hermann-Herder-Strasse 3, 79104 Freiburg, Germany
}
\begin{abstract}
We extend the twin-space formulation of the hierarchical equations of
motion approach in combination with the matrix product state
representation (introduced in J. Chem. Phys. 150, 234102, [2019]) to
nonequilibrium scenarios where the open quantum system is coupled to a
hybrid fermionic and bosonic environment.
The key ideas used in the extension are a reformulation of the
hierarchical equations of motion for the auxiliary density matrices into
a time-dependent Schrödinger-like equation for an augmented
multi-dimensional wave function as well as a tensor decomposition into a
product of low-rank matrices. The new approach facilitates accurate
simulations of non-equilibrium quantum dynamics in larger and more
complex open quantum systems. The performance of the method is
demonstrated for a model of a molecular junction exhibiting
current-induced mode-selective vibrational excitation.
\end{abstract}

\maketitle
\section{Introduction}

Open quantum systems, which are characterized by exchange of particles
or energy with an environment, are widespread in a variety of physical,
chemical, and biological processes,\cite{breuer2002theory,may2008charge} and are relevant to novel
technological developments such as quantum information devices,\cite{loss1998quantum,petersson2012circuit} nanoscale molecular electronics.\cite{Cuevas_2010__p,Galperin_2007_J.Phys.:Condens.Matter_p103201,Bergfield_2013_physicastatussolidib_p2249,Aradhya_2013_Nat.Nanotechnol._p399,Baldea_2016__p,Su_2016_Nat.Rev.Mater._p16002,Thoss_2018_J.Chem.Phys._p30901,Evers_2020_Rev.Mod.Phys._p35001} An accurate theoretical description of open quantum systems has been a long-standing challenge, in particular in cases where the system itself represents a strongly-coupled many-body system which in turn interacts with multiple environments of different type. While approximate methods have provided fundamental insight into non-equilibrium quantum dynamics in open quantum systems,\cite{Cizek_Phys.Rev.B_2004_p125406,Mitra_2004_Phys.Rev.B_p245302,Galperin_2006_Phys.Rev.B_p45314,Timm_2008_Phys.Rev.B_p195416,Hartle_Phys.Rev.Letl._2009_p146801,Lue_2012_Phys.Rev.B_p245444,Thingna_2013_Phys.Rev.E_p52127,Dou_2018_J.Chem.Phys._p102317} it is often crucial to take non-Markovian and non-perturbative effects into account. To this end, several numerically exact techniques have been developed and applied.\cite{shao2004decoupling,Stockburger_PhysRevLett.88.170407,makri1995tensor,suess2014hierarchy,Simine_2012_PhysicalChemistryChemicalPhysics_p13820,Muehlbacher_PhysRevLett.100.176403,Werner_PhysRevB.79.035320,Anders_PhysRevB.74.245113,Cohen_PhysRevLett.115.266802,prior2010efficient,ke2016hierarchy,hsieh2018unified,wang2003multilayer,Wang_2013_J.Phys.Chem.A_p7431,Wang_2016_J.Chem.Phys._p164105,Tanimura_2006_J.Phys.Soc.Jpn._p82001,Xu_2007_Phys.Rev.E_p31107,Shi_2009_J.Chem.Phys._p84105,Ye_2016_WIREsComputMolSci_p608} A promising method in this respect is the hierarchical equations of motion (HEOM) approach. \cite{Tanimura_2020_J.Chem.Phys._p20901}

The HEOM approach was originally proposed by Tanimura and Kubo to study relaxation dynamics of molecular systems subjected to a Gaussian bosonic environment at high-temperature.\cite{Tanimura_1989_J.Phys.Soc.Jpn._p101,tanimura1990nonperturbative} Later on, the method was extended to explore charge transport in quantum dots\cite{Jin_2008_J.Chem.Phys._p234703,Zheng_2008_NewJ.Phys._p93016,PhysRevLett.109.266403,Zheng_2013_Phys.Rev.Lett._p86601} and single-molecule junctions.\cite{Schinabeck_2016_Phys.Rev.B_p201407,Schinabeck_2018_Phys.Rev.B_p235429,Erpenbeck_2020_Phys.Rev.B_p195421} A variety of advanced schemes have been put forward towards broadening the range of applicability, improving the computational efficiency, and removing  numerical issues. For a comprehensive review, we refer the reader to Ref. \onlinecite{Tanimura_2020_J.Chem.Phys._p20901} and the literature therein.  Even with these advances, the method is limited to relatively small model systems, especially when the coupling to the environment is strong, mainly due to the factorial or exponential scaling with respect to the system and effective environmental DoFs. Recently, Shi and coworkers\cite{Shi_J.Chem.Phys._2018_p174102,Yan_J.Chem.Phys._2020_p204109,Yan_J.Chem.Phys._2021_p194104} as well as Borrelli and Gelin,\cite{Borrelli_J.Chem.Phys._2019_p234102,Borrelli_WIREsComputMolSci_2021_pe1539} have separately suggested that the HEOM approach can be combined with the matrix product state (MPS) formulation, also called tensor train (TT) approach. The MPS formulation 
 is an extremely powerful and versatile tool to study quantum many-body physics, in particular for 
 one-dimensional systems
 with low or moderate entanglements.\cite{White_Phys.Rev.Lett._1992_p2863,White_Phys.Rev.B_1993_p10345}\cite{Fannes_Commun.Math.Phys._1992_p443490,Oestlund_Phys.Rev.Lett._1995_p3537,Verstraete_Adv.Phys._2008_p143224,Schollwoeck_Rev.Mod.Phys._2005_p259,Cirac_J.Phys.A_2009_p504004,
McCulloch_J.Stat.Mech.TheoryExp._2007_pP10014,Schollwoeck_Ann.Phys.NY_2011_p96192,GarciaRipoll_NewJ.Phys._2006_p305,Paeckel_Ann.Phys.NY_2019_p167998} MPS has also been applied to 
open quantum system dynamics, in particular 
in connection with other approaches based on a reduced system dynamics description, such as quasi-adiabatic path-integral approach,\cite{Strathearn_NatCommun_2018_p19,Ye_J.Chem.Phys._2021_p044104,Bose_arXivpreprintarXiv2106.14934_2021_p} and hierarchy of pure state.\cite{Flannigan_arXivpreprintarXiv2108.06224_2021_p,Gao_arXivpreprintarXiv2109.06393_2021_p} 
The MPS representation, as a wave function approach, can be considered as a special case of the  multi-layer multi-configurational time-dependent Hartree (ML-MCTDH) method,\cite{wang2003multilayer,vendrell2011multilayer,manthe2008multilayer,meyer2009multidimensional,Larsson_J.Chem.Phys._2019_p204102,Mainali_J.Chem.Phys._2021_p174106} which has also been applied to a variety of prototype open quantum systems.\cite{Wang_J.Phys.Chem.A_2015_p79517965,Manthe_J.Chem.Phys._2015_p244109,Manthe_2017_J.Chem.Phys._p64117,Wang_2018_Chem.Phys._p13}

While it has been demonstrated that the combination of the HEOM approach and tensor train decomposition is a promising way forward, Borrelli\cite{Borrelli_J.Chem.Phys._2019_p234102,Borrelli_WIREsComputMolSci_2021_pe1539} further pointed out that it is instrumental to reformulate the reduced system in twin space, which renders the method more flexible as it allows for the tensor train decomposition in the central system DoFs. In this paper, we extend the work in Ref. \onlinecite{Borrelli_J.Chem.Phys._2019_p234102}  further to handle more complex cases, where the system is in contact with a hybrid fermionic and bosonic environment. The study of these systems is usually prohibitively expensive or even impossible by the conventional HEOM method.

The rest of the paper is organized as follows. We start with an open quantum system model where the environment consists of macroscopic fermionic reservoirs and bosonic baths, before moving on to introduce the conventional HEOM method in Sec. II. Then, we reformulate a hierarchical set of equations for auxiliary density matrices into a time-dependent Schr\"odinger-like equation for an extended wave function, which can then be decomposed in the MPS format and propagated using a time-dependent variational principle scheme.  We provide a benchmark example for a simple electronic two-level model and study the bias-controlled mode-selective vibrational excitation in an asymmetric molecular junction in
Sec. III, and finally conclude and suggest directions for future research in Sec. IV. In all calculations, we use natural units: $\hbar=k_B=e=1$.

\section{Method}
In the language of open quantum system theory, the whole system is divided into the system of interest and its environment. The Hamiltonian is given by 
\begin{equation}
    H=H_{\rm{s}}+H_{\rm{env}}+H_{\rm{s-env}} + H_{\rm{ren}},
\end{equation}
where $H_{\rm{s}}$ and $H_{\rm{env}}$ denote the system and environmental parts, respectively, $H_{\rm{s-env}}$ their coupling, and $H_{\rm{ren}}$ is a renormalization term. 

To be specific, we consider a generic model in a molecular junction setting, where the molecular system is described by a $D$-dimensional Hilbert space, consisting of several electronic levels and a set of vibrational modes. The environment comprises of multiple independent fermionic and bosonic reservoirs, $H_{\rm{env}}=H_{\rm{f}}+H_{\rm{b}}$. Typically, the molecule is connected to two or three macroscopic leads, which can be modelled as a manifold of non-interacting electrons and the corresponding Hamiltonian is given by 
 \begin{equation}
H_{\rm{f}}=\sum_{\alpha}\sum_{k}\epsilon_{\alpha k} c_{\alpha k}^{+}c_{\alpha k}^-,
 \end{equation}
where $c_{\alpha k}^{+}$ $(c^-_{\alpha k})$ denotes the creation (annihilation) operator for an electron in the $k$-th state of lead $\alpha$ with the corresponding energy $\epsilon_{\alpha k}$. 
The lattice motion of the leads and the solvent DoFs constitute the bosonic baths, which can be modelled as phonon baths of harmonic oscillators, 
\begin{equation}
    H_{\rm{b}}=\sum_{\theta }\sum_{k}\omega_{\theta k} a_{\theta k}^{+}a^-_{\theta k},
\end{equation}
where $a_{\theta k}^{\dagger}$ and $a^-_{\theta k}$ are the creation and annihilation operator, respectively, for the $k$-th phonon mode in bath $\theta$ with the frequency $\omega_{\theta k}$.

For the sake of simplicity, we assume that the molecular electronic levels are exclusively coupled to the leads, while the vibrational modes are coupled to the phonon baths, and the interaction Hamiltonian reads
\begin{equation}
\label{interaction_hamiltonian}
H_{\rm{s-env}}=\sum_{i \alpha k }\left(\nu_{i \alpha k} c_{\alpha k}^{+}d^-_{i}+\nu^*_{i \alpha k} d^+_{i}c_{\alpha k}^{-}\right)
+\sum_{j\theta k} \chi_{j\theta k} x_j (a_{\theta k}^{+}+a^-_{\theta k}).
 \end{equation}
Here, $\nu_{i\alpha k}$ specifies the coupling strength between the $i$-th molecular electronic level (with creation and annihilation operators $d_i^+$ and $d_i^-$, respectively) and the $k$-th state in lead $\alpha$. The interaction Hamiltonian also contains a bilinear coupling between the $j$-th vibrational mode ($x_j$ is the position operator) and the $k$-th phonon mode in bath $\theta$ where $\chi_{j\theta k}$ determines the coupling strength. Note that the method presented below can be readily extended to more complicated interacting cases. 

The renormalization term
\begin{equation}
    H_{\rm{ren}}=\sum_{j \theta }\left(\sum_{k}\frac{ \chi_{j\theta k}^2  }{\omega_{\theta k}^2} \right) x_j^2= \sum_{j \theta }\frac{\Lambda_{j \theta } }{2}x_j^2
\end{equation}
is introduced to counteract the artificial change of the system potential due to the coupling to the phonon baths. 

Furthermore, we assume that initially, at $t=0$, the environments are disentangled from the system and prepared at their own thermal equilibrium with temperature $T$ (or inverse temperature $\beta=1/T$). After integrating out the environmental DoFs, and due to their Gaussian statistical properties, the influence of the coupling given above in \Eq{interaction_hamiltonian} on the system dynamics is exclusively encoded in
the thermal equilibrium correlation functions:
\begin{equation}
C_{i\alpha}^{\sigma}(t)
=\frac{1}{2\pi}\int_{-\infty}^{\infty} e^{i \sigma\epsilon t}\Gamma_{i\alpha }(\epsilon)f_{\alpha}^{\sigma}(\epsilon)\mathrm{d}\epsilon
\end{equation}
and
\begin{equation}
    W_{j\theta}(t) = \frac{1}{2\pi} \int_{-\infty}^{\infty} e^{-i\omega t} J_{j\theta}(\omega)f_B(\omega)\mathrm{d}\omega.
\end{equation}
In the above formulae, $\sigma$ can be $+$ or $-$. The Fermi-Dirac distribution $f_{\alpha}^{\sigma}(\epsilon)=\frac{1}{e^{\sigma\beta (\epsilon- \mu_{\alpha})}+1}$, denotes the distribution of electrons/holes ($\sigma=+/-$) in lead $\alpha$ with chemical potential $\mu_{\alpha}$. The Bose-Einstein distribution function of phonons is described by
$f_B( \omega ) = \frac{1}{e^{\beta \omega}-1}$.
$\Gamma_{i\alpha}(\epsilon)$ and $ J_{j\theta}(\omega)$ are the so-called spectral density functions and   are defined as
\begin{equation}
\Gamma_{i\alpha}(\epsilon)=2\pi \sum_{k} |\nu_{i\alpha k}|^2 \delta (\epsilon-\epsilon_{\alpha k}),
\end{equation}
and
\begin{equation}
    J_{j\theta}(\omega)=2\pi \sum_k \frac{\chi_{j\theta k}^2}{\omega_{\theta k}} \delta(\omega - \omega_{\theta k}).
\end{equation}
In  this work, we adopt the wide-band approximation of the leads, i.e., 
\begin{equation}
    \Gamma_{i\alpha}(\epsilon)=\Delta^2_{i\alpha},
\end{equation}
where $\Delta_{i \alpha }$ is a constant. Additionally, we assume that the spectral density function of the bosonic baths takes a Lorentzian form,
\begin{equation}
\label{lorentzian}
   J_{j\theta}(\omega)=
2\lambda^2_{j\theta} \frac{\omega \Omega_{\theta}}{\omega^2 + \Omega_{\theta}^2} ,
\end{equation} 
with two characteristic parameters, the coupling strength $\lambda_{j\theta}$ and the cut-off frequency $\Omega_{\theta}$, resulting in
\begin{equation}
    \Lambda_{j\theta }=\frac{1}{\pi}\int_{-\infty}^{\infty} \frac{J_{j\theta}(\omega)}{\omega} \mathrm{d}\omega =2\lambda_{j\theta}^2 .
\end{equation} 
By employing the Pad\'e pole decomposition scheme of the Fermi-Dirac and Bose-Einstein distribution functions, $f_{\alpha}^{\sigma}(\epsilon)$ and $f_B(\omega)$ (with the pole numbers $P_{\rm{f}}$ and $P_{\rm{b}}$, respectively), the correlation function can be expanded as a sum of exponential functions,\cite{Hu_2010_J.Chem.Phys._p101106,Hu_2011_J.Chem.Phys._p244106,Cui_2019_J.Chem.Phys._p24110,Abe_Phys.Rev.B_2003_p235411}  with
\begin{equation}
\label{corr_f_exp}
    C_{i\alpha}^{\sigma}(t) \simeq \pi \delta(t) + \sum_{p=1}^{P_{\rm{f}}}\Delta^2_{i\alpha}\eta_{\alpha p}e^{-\gamma_{\alpha p}^{\sigma}t}
\end{equation}
and
\begin{eqnarray}
\label{corr_b_exp}
    W_{j\theta}(t)\simeq \sum_{p=0}^{P_{\rm{b}}}\lambda^2_{j \theta}\eta_{\theta p}e^{-\gamma_{\theta p}t} +\sum_{p=P_{\rm{b}}+1}^{\infty}\frac{\lambda^2_{j \theta}\eta_{\theta p}}{\gamma_{\theta p}}  \delta(t).
\end{eqnarray}
The explicit expressions of the coefficients $\{\eta\}$ and exponents $\{\gamma\}$ can be found in the supplementary material. We will also employ the Markovian approximation for high-frequency components from \Eq{corr_b_exp}. For sufficiently large $\gamma_{\theta p}$, the rapidly decaying exponential terms are replaced by delta functions. This closure has been reported to ease the stability and positivity issues of the HEOM approach at low temperatures. \cite{Ishizaki_2005_J.Phys.Soc.Jpn._p3131} 

By taking advantage of the self-similarity of the exponential functions with respect to a time-derivative, one can introduce a group of auxiliary density operators (ADOs) $\{\rho^{\bm{n,m}}(t)\}$ and formulate a hierarchical set of equations of motion,\cite{Jin_2008_J.Chem.Phys._p234703,Shi_2009_J.Chem.Phys._p84105,Hsieh_2018_J.Chem.Phys._p14103,Xu_2019_J.Chem.Phys._p44109,Baetge_Phys.Rev.B_2021_p235413,Ke_J.Chem.Phys._2021_p234702}
\begin{widetext}
\begin{eqnarray}
\label{heom}
\frac{d \rho^{\bm{n,m}}(t) }{dt} 
&=&- i\left[H_{\rm{s}}+H_{\rm{ren}},\rho^{\bm{n,m}}(t)\right] 
+{\sum_{k=1}^K n_{k}\gamma^{\sigma_k}_{\alpha_k p_k}  \rho^{\bm{n,m} }  }(t) 
+{\sum_{l=1}^L m_{l}\gamma_{\theta_l p_l}  \rho^{\bm{n,m} }  }(t)
\\
&& 
-\sum_{i\alpha \sigma} \frac{\Delta^2_{i\alpha}}{4} 
\left[d_{i}^{\bar{\sigma}}, \left[d_{i}^{\sigma}, \rho^{\bm{n},\bm{m}}(t)\right]_{(-)^{||\bm{n}||+1}}\right]_{(-)^{||\bm{n}||+1}}
-\sum_{j\theta} \lambda^2_{j \theta}\Lambda_{\theta}
\left[x_j, \left[x_j, \rho^{\bm{n},\bm{m}}(t)\right]_{-}\right]_{-}
\nonumber
\\
 &&+i\sum_{k=1}^{K} (-1)^{\sum_{j<k}n_{j}}  \sqrt{1-n_k} \Delta_{i_k \alpha_k}\left(  d_{i_k}^{\bar{\sigma}_k} \rho^{\bm{n}+{\bm{1}}_k,\bm{m}}(t)+
   (-1)^{||\bm{n}||+1}\rho^{\bm{n}+\bm{1}_k,\bm{m}} (t)d^{\bar{\sigma}_k}_{i_k} \right)\nonumber \\
&& +i\sum_{k=1}^{K} (-1)^{\sum_{j<k}n_{j}}  \sqrt{n_k} \Delta_{i_k \alpha_k}\left(\eta_{\alpha_k p_k} d^{\sigma_k}_{i_k} \rho^{\bm{n}-\bm{1}_k,\bm{m}}(t)  
{ -(-1)^{||\bm{n}||-1}}
 \rho^{\bm{n}-\bm{1}_k,\bm{m}}(t)  \eta_{\alpha_kp_k}^{*} d^{\sigma_k}_{i_k}\right)  \nonumber \\
 &&+i\sum_{l=1}^{L} \sqrt{m_l+1} \lambda_{j_l \theta_l}\left( x_{j_l} \rho^{\bm{n},\bm{m}+{\bm{1}}_l}(t)-
   \rho^{\bm{n},\bm{m}+\bm{1}_l} (t)x_{j_l}\right)\nonumber \\
&& +i\sum_{l=1}^{L} \sqrt{m_l} \lambda_{j_l \theta_l}\left(\eta_{\theta_l p_l} x_{j_l} \rho^{\bm{n},\bm{m}-\bm{1}_l}(t)  
-\rho^{\bm{n},\bm{m}-\bm{1}_l}(t)  \eta_{\theta_l p_l}^{*} x_{j_l}\right).  \nonumber 
\end{eqnarray}
\end{widetext}

In the HEOM, \Eq{heom}, we have introduced several notations. The expression
\begin{equation}
\Lambda_{\theta}=\frac{2}{\beta \Omega_{\theta}}-\cot(\beta\Omega_{\theta}/2)-\sum_{p=1}\frac{\eta_{\theta p}}{\gamma_{\theta p}}   
\end{equation} 
originates from the Markovian approximation.  
$[A,B]_{(-)^n}$ denotes the commutator (anti-commutator) betwen $A$ and $B$, when $n$ is an odd (even) number.
The first bold index $\bm{n}$ in the superscript is given by
\begin{equation}
  \bm{n}=(n_{1}, n_{2},\cdots, n_{k}, \cdots, n_{K}),  
\end{equation} 
where $k$ runs from 1 to $K=2 D_e N_{\alpha}P_{\rm{f}}$. Here $D_e$ denotes the number of molecular electronic levels, $N_{\alpha}$ the number of leads, and $P_{\rm{f}}$ the number of fermionic Pad\'e poles, respectively. The elements 
$n_k$ can be either $0$ or $1$, and their summation equals to the norm $||\bm{n}||=\sum_{k=1}^K n_k$. 
We can consider $n_k$ as the occupation number of a \enquote{virtual} electronic level specified by four indices $i_k$, $\alpha_k$, $p_k$, and $\sigma_k$. Here, $i_k \in \{1, \cdots, D_e\}$ specifies the molecular electronic DoFs,  $\alpha_k$ is the lead index, $p_k \in \{1,\cdots, P_{\rm{f}}\}$ is the index of fermionic Pad\'e poles, and $\sigma_k \in \{+,-\}$ is the binary sign with conjugation value $\bar{\sigma}_k=-\sigma_k$. The notation
$\bm{n}\pm \bm{1}_k$ is given as 
\begin{equation}
  \bm{n}\pm \bm{1}_k = (n_{1}, n_{2},\cdots, 1-n_{k}, \cdots, n_{K}). 
\end{equation}
The other bold index $\bm{m}$  accounts for the phonon baths and is given by
\begin{equation}
  \bm{m}=(m_{1}, m_{2},\cdots, m_{l}, \cdots, m_{L}),  
\end{equation} 
where $l$ runs over 1 to $L=D_{\rm{vib} } N_{\theta} (P_{\rm{b}}+1)$, with  $D_{\rm{vib} }$, $N_{\theta}$ and $P_{\rm{b}}$ being the number of system vibrational modes, phonon baths and bosonic Pad\'e poles, respectively. Similarly to the electron reservoir index, the index $m_l \in \{0, 1, \cdots \}$ is a non-negative integer number, which can be interpreted as the occupation number of a \enquote{virtual} bosonic mode specified by three indices: $j_l$, $\theta_l$, and $p_l$. Analogously to before, $j_l \in \{1, \cdots, D_{\rm{vib} }\}$ specifies the system vibrational mode,  $\theta_l$ indicates which phonon bath it represents, and $p_l \in \{0,1,\cdots, P_{\rm{b}}\}$ is the index of bosonic Pad\'e poles. An increase or decrease of one vibrational quantum at the $l$-th mode corresponds to
\begin{equation}
  \bm{m\pm 1}_l=(m_{1}, m_{2},\cdots, m_{l}\pm 1, \cdots, m_{L}).  
\end{equation} 

The ADO $\rho^{\bm{n,m}}$ (for a given $\bm{n}$ and $\bm{m}$) is normally represented in Hilbert space as a matrix. However, we will see that it is useful for the tensor train approach to reformulate it into a "vector" in  twin-space.\cite{Borrelli_J.Chem.Phys._2019_p234102,Borrelli_WIREsComputMolSci_2021_pe1539} This process is also known as purification in quantum computing.\cite{Nielsen__2010_p,Verstraete_Phys.Rev.Lett._2004_p207204,Feiguin_Phys.Rev.B_2005_p220401}
The twin-space formulation was introduced in thermo-field theory.\cite{Schmutz_ZeitschriftfurPhysikBCondensedMatter_1978_p97106,Suzuki_J.Phys.Soc.Jpn._1985_p44834485,Arimitsu_Prog.Theor.Phys._1987_p3252,Suzuki_Int.J.Mod.Phys.B_1991_p18211842}
It is obtained by constructing a set of ancillary states $\{|\tilde{s}_i\rangle\}$ in one-to-one correspondence with the physical states $\{|s_i\rangle\}$ for the $i$-th system DoF. In other word, the ancillas (the fictitious DoFs) form a copy of the original system, doubling the size of the system subspace. Within the twin-space formulation, the system subspace is spanned by a new orthogonal complete basis set $\{\bigotimes_{i=1}^D |s_i\rangle |\tilde{s}_i\rangle\}$. The unit vector in twin space is 
\begin{equation}
\label{unit_vector}
    |\mathbbm{1}\rrangle = \bigotimes_{i=1}^D \sum_{s_i =\tilde{s}_i}|s_i\rangle |\tilde{s}_i\rangle,
\end{equation}
and the auxiliary density matrix, 
\begin{equation}
    \rho^{\bm{n,m}}(t) =\sum_{\begin{subarray}{l} s_1\cdots s_D\\ \tilde{s}_1\cdots \tilde{s}_D\end{subarray}}
    C^{\bm{n,m}}_{s_1\cdots s_D \tilde{s}_1\cdots \tilde{s}_D}(t) 
    |s_1\cdots s_D \rangle \langle \tilde{s}_D \cdots  \tilde{s}_1|,
\end{equation}
is transformed into
\begin{equation}
\label{reduced_density matrix}
    |\rho^{\bm{n,m}}(t) \rrangle= \sum_{\begin{subarray}{l} s_1\cdots s_D \\ \tilde{s}_1 \cdots \tilde{s}_D \end{subarray}}
    C^{\bm{n,m}}_{ s_1 \tilde{s}_1 \cdots s_D \tilde{s}_D}(t) 
    |s_1\rangle \otimes |\tilde{s}_1\rangle  \otimes \cdots \otimes|s_D\rangle \otimes |\tilde{s}_D\rangle
    .
\end{equation}
Correspondingly, we introduce two special pairs of super-operators in twin-space, $\hat{d}^{\pm}_i$ and $\tilde{d}^{\pm}_i$, as well as $\hat{x}_j$ and $\tilde{x}_j$, acting on $|\rho\rrangle$ as
\begin{subequations}
\begin{align}
     \hat{d}^{\pm}_i  |\rho\rrangle &= d^{\pm}_i \otimes \mathbbm{1}_i^e |\rho\rrangle \coloneq d^{\pm}_i \rho ,\\
    \tilde{d}^{\pm}_i  |\rho\rrangle &= \mathbbm{1}_i^e \otimes d^{\mp}_i |\rho\rrangle \coloneq \rho d^{\mp}_i      ,\\
    \hat{x}_j  |\rho\rrangle &= x_j \otimes \mathbbm{1}_j^{ \rm{vib}} |\rho\rrangle \coloneq x_j \rho ,\\
    \tilde{x}_j  |\rho\rrangle &= \mathbbm{1}_j^{\rm{vib} } \otimes x_j |\rho\rrangle \coloneq \rho x_j .
\end{align}
\end{subequations}
The super-operators with a hat (\enquote{$\,\, \hat{}\,\,$}) act on the physical DoFs, while those with a tilde (\enquote{$\,\,\tilde{}\,\,$}) act on ancilla DoFs.

We should point out that the twin-space is not an alias for the Liouville space for many-body systems. But there is a one-to-one mapping between the states in the two formalisms induced by the identity vector.\cite{Schmutz_ZeitschriftfurPhysikBCondensedMatter_1978_p97106} In the Liouville space, one puts all the physical DoFs in one block and the ancillas in the other, instead of pairing each physical DoF with its ancilla in an alternative manner. A visualization of this explanation and more discussion concerning their difference are provided in the supplementary material.

As implied above, we can assume that the bold index $\bm{n}$ corresponds to a Fock state of fermions.
For the single-level vacuum state $|0_k\rangle$, the creation operator $c^{<,+}_{k}$ will fill the state with a fermion, $c^{<,+}_{k}|0_k\rangle=|1_k\rangle$.
The state $|\bm{n}\rangle$ is obtained by acting a sequence of creation operators $c^{<,+}_k$ on the vacuum state, 
\begin{equation}
    |\bm{n}\rangle =|n_{1}\cdots n_{k} \cdots n_{K}\rangle 
    =(c^{<,+} _{1} )^{n_1}  \cdots 
    (c^{<,+}_{k} )^{n_k}\cdots    (c^{<,+}_{K} )^{n_{K}}
    |\underbrace{0\cdots 0}_{K}\rangle.
\end{equation}
 Applying the fermion-like creation and annihilation operators  $c^{<,+}_{k}$, $c^{<,-}_{k}$ on the Fock state $|\bm{n}\rangle$ yields
\begin{subequations}
\begin{align}
c^{<,+}_{k}& |\bm{n}\rangle=(-1)^{\sum_{j<k}n_j}\sqrt{1-n_k}|\bm{n}+\bm{1}_k\rangle,\\
c^{<,-}_{k}& |\bm{n}\rangle=(-1)^{\sum_{j<k}n_j}\sqrt{n_k}|\bm{n}-\bm{1}_k\rangle,\\
c^{<,+}_{k}& c^{<,-}_{k}|\bm{n}\rangle=n_k|\bm{n}\rangle.
\end{align}
\end{subequations}
It is also necessary to introduce another set of fermion-like creation and annihilation operators $c^{>,+}_{k}$, $c^{>,-}_{k}$, and a special operator $I^{>}$, which act on the Fock state $|\bm{n}\rangle$ as
\begin{subequations}
\begin{align}
 c^{>,+}_{k}&|\bm{n}\rangle=(-1)^{\sum_{j>k}n_j}\sqrt{1-n_k}|\bm{n}+\bm{1}_k\rangle,\\
 c^{>,-}_{k}& |\bm{n}\rangle=(-1)^{\sum_{j>k}n_j}\sqrt{n_k}|\bm{n}-\bm{1}_k\rangle, \\
 I^{>}&|\bm{n}\rangle=(-1)^{\sum_{j=1}^{K}n_j}|\bm{n}\rangle.
 \end{align}
\end{subequations}

We similarly assume that $|\bm{m}\rangle$ is a Fock state of virtual bosons, which though does not correspond to any specific environmental state, and it is generated by
\begin{equation}
    |\bm{m}\rangle =|m_{1}\cdots m_{l} \cdots m_{L}\rangle 
    =(b^+_{1} )^{m_1}  \cdots 
    (b^+_{l} )^{m_l}\cdots    (b^+_{L} )^{m_L}
    |\underbrace{0\cdots 0}_L\rangle.
\end{equation}
Here, the bosonic creation $b_l^+$ and annihilation $b_l^-$ operators are introduced, and they are applied upon the state $|\bm{m}\rangle$ to yield
\begin{subequations}
\begin{align}
    b^+_{l}&|\bm{m}\rangle = \sqrt{m_l+1}|\bm{m}+\bm{1}_l\rangle,\\
    b^-_{l}&|\bm{m}\rangle = \sqrt{m_l}|\bm{m}-\bm{1}_l\rangle,\\
    b^+_l &b^-_l|\bm{m}\rangle = m_l |\bm{m}\rangle.
\end{align}
\end{subequations}

\begin{figure*}
\centering
	\begin{minipage}[c]{0.8\textwidth}		
			\raggedright a) \\
			\includegraphics[width=\textwidth]{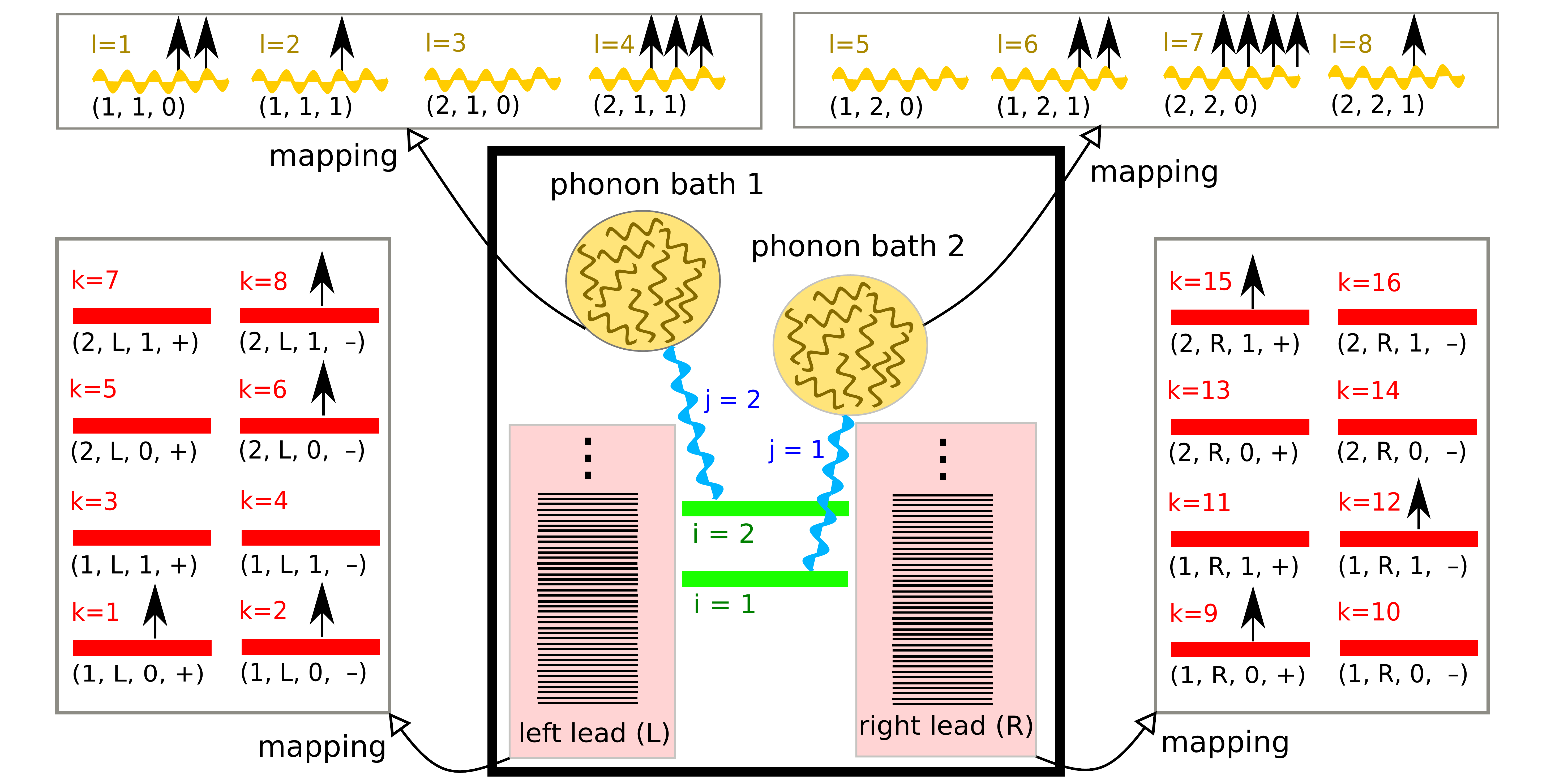}\\
			\raggedright b) \\
			\includegraphics[width=\textwidth]{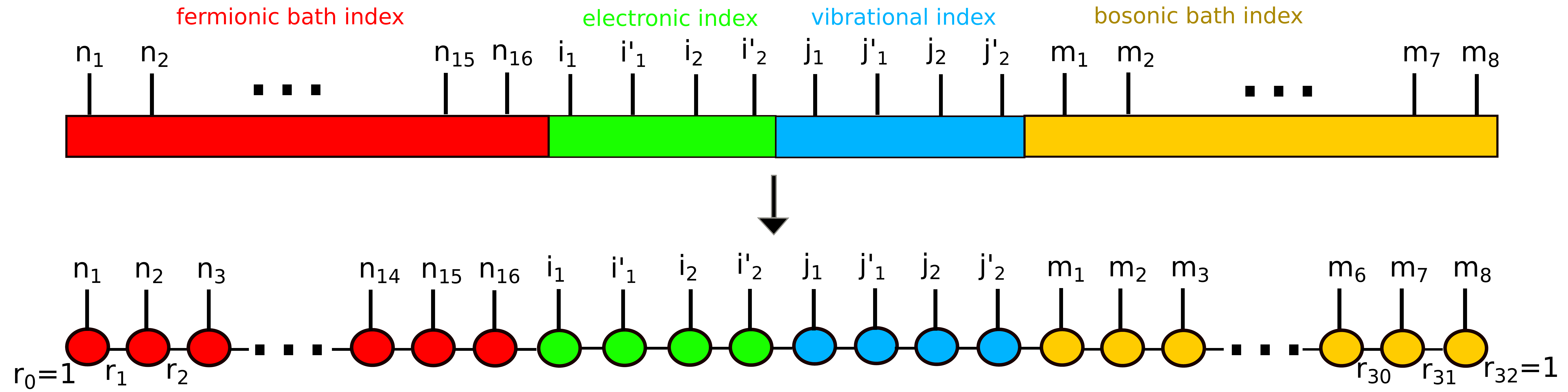}\\
	\end{minipage}
 \caption{(a) Schematic illustration of mapping the continuum of lead states into a few discrete virtual electronic levels (red bars), and phonon baths with infinite harmonic oscillators into a finite number of harmonic oscillators (yellow wavy lines), within the HEOM approach. As an example, we display in the black thick rectangle a system comprising two electronic levels (green bars) coupled to two leads and two vibrational modes (blue wavy lines), with the latter coupled to its own phonon bath. For the purpose of illustration, we include only two Pad\'e poles for the spectral decomposition of the Fermi-Dirac/Bose-Einstein function: $D_e=2, D_{\rm{vib}}=2, N_{\alpha} = 2, N_{\theta}=2, P_{\rm{f}} =2, P_{\rm{b}}=2$. This results in a fictitious extended system, where two leads are mapped into 16 effective virtual electronic levels and two phonon baths into 8 effective damped harmonic oscillators. We have also included an example occupancy of the discrete environmental states, which then correspond to an auxiliary density matrix $\rho^{\bm{n,m}}$ with $\bm{n}=(n_1,\cdots, n_{16})=(1,1,0,0,0,1,0,1,1,0,0,1,0,0,1,0)$, and $\bm{m}=(m_1,\cdots,m_8)=(2,1,0,3,0,2,4,1)$. The notations below the red bar correspond to the four indices of the virtual electronic levels, $(i_k$, $\alpha_k$, $p_k$, $\sigma_k)$, and those below the yellow wavy lines to the virtual bosonic levels, $(j_l$, $\theta_l$, $p_l)$. (b) Schematic illustration of recasting the above extended system $|\Psi\rangle$ as a rank-M tensor into a one-dimensional architecture and then decomposing it further into tensor train format, which allows a one-to-one match between the different DoFs (including physical,  tilde, and virtual DoFs) and the core tensors. The red and orange circles appearing at the head and end block of the chain represent the tensors corresponding to the virtual fermionic states of the leads and virtual bosonic states of the phonon baths, respectively. The green and blue circles correspond to the system tensors for electronic and vibrational degrees of freedom, respectively, and every site is adjacent to its ancilla. The vertical dangling legs correspond to the physical index and the connected legs mean the contraction between the tensors.}
 \label{fig1}
\end{figure*}

At this point, the auxiliary density matrices $\{\rho^{\bm{n},\bm{m}}\}$ can be recast as an augmented $M(=2D+K+L)$-dimensional wave function (from now on the time argument is omitted):
\begin{widetext}
\begin{equation}
\label{extended_wavefunction}
    |\Psi \rangle=|\bm{n} \rangle \otimes |\rho^{\bm{n,m}} \rrangle \otimes |\bm{m} \rangle 
    =\sum_{\begin{subarray}{c} n_1\cdots n_K  m_1\cdots m_L \\
    s_1 \tilde{s}_1\cdots s_D \tilde{s}_D \\  \end{subarray}}
    C_{ s_1\tilde{s}_1 \cdots s_D \tilde{s}_D}^{n_1\cdots n_K,m_1 \cdots m_L}
    |n_1 \cdots n_K\rangle  
   \otimes|s_1\tilde{s}_1 \cdots s_D\tilde{s}_D\rangle \otimes  |m_1 \cdots m_L\rangle. 
\end{equation}
It can be interpreted as an extended system where the leads are mapped into $K$ effective virtual electronic levels, and the phonon baths into $L$ effective fictitious bosonic modes, as schematically shown in \Fig{fig1} (a). If all these effective virtual states are unpopulated, it reproduces the reduced system dynamics, corresponding to the reduced density matrix of the system,
\begin{equation}
    \rho_s=|\Psi_s \rrangle =|\bm{0} \rangle \otimes |\rho^{\bm{0,0}} \rrangle \otimes|\bm{0} \rangle  
    =\sum_{\begin{subarray}{c}
    s_1 \tilde{s}_1 \cdots s_D  \tilde{s}_D \end{subarray}}
    C_{ s_1 \tilde{s}_1 \cdots s_D \tilde{s}_D}^{0 \cdots 0,0\cdots 0}
    |\underbrace{0 \cdots 0}_{K}\rangle \otimes
    |s_1\tilde{s}_1\cdots s_D \tilde{s}_D \rangle 
    \otimes|\underbrace{0 \cdots 0}_L\rangle . 
\end{equation}
\end{widetext}

We can now rewrite the hierarchical equations of motion in \Eq{heom} into a time-dependent Schr\"odinger-like equation for $|\Psi\rangle$,
\begin{equation}
\label{final_equation}
i\frac{d |\Psi\rangle }{dt} =\mathbbm{H}
 |\Psi\rangle,
\end{equation}
where the super-Hamiltonian in this further enlarged space is written as
\begin{eqnarray}
\label{Hamiltonian}
\mathbbm{H} &=&\hat{H}_s + \hat{H}_{\rm{ren}} -\tilde{H}_s  - \tilde{H}_{\rm{ren}}
-i\sum_{k=1}^{K} \gamma^{\sigma_k}_{\alpha_k p_k} c^{<,+}_{k}   c^{<,-}_{k}  \\
&&
-i\sum_{l=1}^{L} \gamma_{\theta_l p_l} b^{+}_{l}   b^{-}_{l}
-\sum_{i \alpha \sigma} \frac{\Delta^2_{i\alpha}}{4} 
(\hat{d}_{i}^{\bar{\sigma}} - I^{>} \tilde{d}_{i}^{\bar{\sigma}})\cdot (\hat{d}_{i}^{\sigma}-I^{>}\tilde{d}_{i}^{\sigma})  \nonumber \\
&&
 -\sum_{j\theta} \lambda^2_{j\theta} \Lambda_{\theta}
( \hat{x}_{j} - \tilde{x}_{j} )^2
 \nonumber 
 -\sum_{k=1}^{K}  \Delta_{i_k \alpha_k} \left( c^{<,-}_{k} \hat{d}^{\bar{\sigma}_k}_{i_k} -  c^{>,-}_{k} \tilde{d}_{i_k}^{\bar{\sigma}_k}\right)  \nonumber \\
 &&
-\sum_{k=1}^{K}\Delta_{i_k \alpha_k} \left( \eta_{\alpha_k p_k}  c^{<,+}_{k}  \hat{d}^{\sigma_k}_{i_k}  -\eta_{\alpha_kp_k}^{*}  c^{>,+}_{k}  \tilde{d}^{\sigma_k}_{i_k}\right) \nonumber \\
&&
 -\sum_{l=1}^{L}  \lambda_{j_l \theta_l} \left(\hat{x}_{j_l} - \tilde{x}_{j_l} \right)b^{-}_{l} 
-\sum_{l=1}^{L}\lambda_{j_l \theta_l}\left( \eta_{\theta_l p_l} 
\hat{x}_{j_l} -\eta_{\theta_l p_l}^{*} 
 \tilde{x}_{j_l} \right) b^{+}_{l}.  \nonumber 
\end{eqnarray}

\Eq{final_equation} and \Eq{Hamiltonian} are the main result of this work, which in combination with the tensor train decomposition of the wave function $|\Psi\rangle$ presented below, is called the HEOM+TT method.

In general, the vibrational basis set can be infinite. However, bosonic populations typically decrease in  states with a higher quantum number. It is reasonable, therefore, to retain in practice only a finite number of states, $N_{\rm{vib} }$, for the physical vibrational modes and their ancillas, and truncate the bosonic hierarchy to $N_h$ for the virtual environmental phonon modes. Thus, the state vector $|\Psi\rangle$ contains $2^{K+2D_e}\times N_{\rm{vib} }^{2D_{\rm{vib} }}\times N_h^L$ elements. Despite the fact that the sparsity of $|\Psi\rangle$ and $\mathbbm{H}$ is very high\cite{Hou_2015_J.Chem.Phys._p104112,Ye_2016_WIREsComputMolSci_p608} and that one can use specialized algorithms to solve \Eq{final_equation}, it is still intractable to store all nonzero elements once $M$ is large enough. This is the case for large system size or low temperature, which prevents the direct application of the conventional HEOM approach. However, we can take full advantage of the techniques developed for the propagation of multi-dimensional wave functions.  

One efficient approach is to bring $|\Psi\rangle$ into the matrix product state format,\cite{Schollwoeck_Ann.Phys.NY_2011_p96192} which is also called tensor train.  In this method, the time-dependent  high-rank coefficient tensor $C_{ s_1\tilde{s}_1\cdots s_D \tilde{s}_D}^{n_1\cdots n_K,m_1 \cdots m_L}$ is decomposed into a tensor product of low-rank matrices, written explicitly as 
\begin{widetext}
\begin{eqnarray}
      C_{ s_1 \tilde{s}_1 \cdots s_D \tilde{s}_D}^{n_1 \cdots n_K,m_1 \cdots m_L}&=&
        A^{[1]}(n_1) \cdots  A^{[K]}(n_K) 
       A^{[K+1]}(s_1)  \cdots A^{[K+2D]}(\tilde{s}_D)
       A^{[K+2D+1]}(m_1)   \cdots A^{[K+2D+L]}(m_L) \\
               &=& \sum_{r_0 r_1  \cdots r_{K+2D+L}}       A^{[1]}(r_0,n_1,r_1)   \cdots  A^{[K+2D+L]}(r_{K+2D+L-1},m_L, r_{K+2D+L}). \nonumber
\end{eqnarray}
\end{widetext}
The graphical representation of the procedure is illustrated in \Fig{fig1} (b). The rank-3 tensors $A^{[i]}$ are called the cores of the MPS/TT decomposition. For the physical index $n_i$, $A^{[i]}(n_i)$ is an $r_{i-1}\times r_i$ complex-valued matrix. The dimensions $r_i$ are called compression ranks or bond dimensions. Specifically, the first and the last rank are fixed as  $r_0=r_{K+2D+L}=1$, such that the matrices multiply into a scalar. This way, the elements to be kept are vastly reduced to at most $(2K+4D_e+2N_{\rm{vib} }D_{\rm{vib} } +N_hL)r_{\rm{max}}^2$, where $r_{\rm{max}}$ is the maximum value of the ranks.

Similarly to the MPS/TT description of the wave function, the super-Hamiltonian $\mathbbm{H}$ can also be expressed in the matrix product operator (MPO) format as
\begin{equation}
\begin{split}
    \mathbb{H}=&
        X^{[1]}(n_1,n'_1) \cdots  X^{[K]}(n_K,n'_K) \\
        &X^{[K+1]}(s_1, s'_1)\cdots  X^{[K+2D]}(\tilde{s}_D, \tilde{s}'_D) \\
        &X^{[K+2D+1]}(m_1,m'_1) \cdots  X^{[K+2D+L]}(m_L,m'_L)
 \end{split}
\end{equation}
where $X^{[i]}$ are rank-4 tensors and obtained by repeatedly performing a sequence of Kronecker products, standard MPO addition and single value decomposition (SVD) truncation with a prescribed accuracy $\varepsilon$ to control the ranks of tensor train matrices, as elaborated in Refs. \onlinecite{Borrelli_WIREsComputMolSci_2021_pe1539}, \onlinecite{Schollwoeck_Ann.Phys.NY_2011_p96192},  and \onlinecite{Oseledets_SIAMJ.Sci.Comput._2011_p22952317}.

Several methods have been developed to compute the time evolution of the MPS/TT representation, and we refer the reader to Ref. \onlinecite{Paeckel_Ann.Phys.NY_2019_p167998} for a thorough review. In this work, we employ the one-site version of the time-dependent variational principle (TDVP) scheme,\cite{Haegeman_Phys.Rev.B_2016_p165116} which is appealing since it is applicable to arbitrary Hamiltonians in the MPO format.  The method solves the dynamical equations projected onto a manifold $\mathcal{M}_{TT}$, which is the set of MPS/TT with fixed ranks,  using a splitting scheme over the tensor train cores. The resulting equation of motion is written formally as
\begin{equation}
\label{TDVP}
    \frac{d}{dt}|\Psi(A(t))\rangle = -i P_{T(A(t))}\mathbbm{H} | \Psi(A(t))\rangle 
\end{equation}
where A labels all the cores of the MPS/TT representation. The notation $P_{T(A(t))}$ denotes the orthogonal projection into the tangent space of $\mathcal{M}_{TT}$ at $| \Psi(A(t))\rangle$. As such, the solution of \Eq{TDVP} is constrained within the manifold $\mathcal{M}_{TT}$, being the best approximation to the actual wave function. This projection incurs an error, because the true time evolution of wave function $\mathbbm{H} | \Psi(t)\rangle$ can run out of the manifold $\mathcal{M}_{TT}$. The explicit differential equations and their approximation properties are analyzed in Refs \onlinecite{Haegeman_Phys.Rev.B_2016_p165116}, \onlinecite{Lubich_SIAMJ.Numer.Anal._2015_p917941}, \onlinecite{Lubich_SIAMJ.MatrixAnal.Appl._2013_p470494}, and \onlinecite{Haegeman_Phys.Rev.B_2013_p075133}.

It is worth noting that the HEOM given in \Eq{heom}, and, correspondingly, the super-Hamiltonian $\mathbbm{H}$, is not unique. The equation can vary slightly according to the different definitions of ADOs, while producing the same reduced system dynamics. We found that, although in the conventional HEOM approach, different expressions do not lead to significant differences in their numerical performance, in the MPS/TT format, the non-uniqueness property provides great flexibility in optimizing the performance of the method. More details on this aspect and numerical demonstrations can be found in the supplementary material.

\section{Results}
\subsection{Electronic Two-level Model}
In order to demonstrate the applicability and validity of the approach in \Eq{final_equation}, we start by benchmarking our results against the conventional HEOM approach for a simple electronic two-level model. The system Hamiltonian is given by 
\begin{equation}
\label{twolevel}
H_s= \epsilon_{1} d_{1}^+ d_{1}^- 
+\epsilon_{2} d_{2}^+d_{2}^-
+V (d_{1}^{+}d_{2}^-+ d_{2}^{+}d_{1}^-) + Ud_{1}^{+}d_{1}^- d_{2}^{+}d_{2}^- ,\\
 \end{equation}
which in twin space is represented as
\begin{subequations}
\begin{align}
\hat{H}_s= \epsilon_{1} \hat{d}_{1}^+\hat{d}_{1}^- 
+\epsilon_{2} \hat{d}_{2}^+\hat{d}_{2}^-
+V (\hat{d}_{1}^{+}\hat{d}_{2}^-+ \hat{d}_{2}^{+}\hat{d}_{1}^-) + U\hat{d}_{1}^{+}\hat{d}_{1}^- \hat{d}_{2}^{+}\hat{d}_{2}^- ,\\
\tilde{H}_s= \epsilon_{1} \tilde{d}_{1}^+\tilde{d}_{1}^- 
+\epsilon_{2} \tilde{d}_{2}^+\tilde{d}_{2}^-
-V (\tilde{d}_{1}^{+}\tilde{d}_{2}^- + \tilde{d}_{2}^{+}\tilde{d}_{1}^- ) + U\tilde{d}_{1}^{+}\tilde{d}_{1}^- \tilde{d}_{2}^{+}\tilde{d}_{2}^-,
\end{align}
\end{subequations}
 where $\epsilon_{1}$ and  $\epsilon_{2}$ are the on-site energies for the two electronic levels, respectively. $V$ denotes the transfer coupling between two states and $U$ the Coulomb interaction. The system is coupled to two leads, labeled by $L$ and $R$. The coupling strengths of two electronic levels to two leads are the same with the value $\Gamma$. The coupling to the phonon baths is neglected for the moment.
 
 \begin{figure*}
	\begin{minipage}[c]{0.31\textwidth}		
			\raggedright a) $\Gamma=\Delta^2=0.01$ eV \\
			\includegraphics[width=\textwidth]{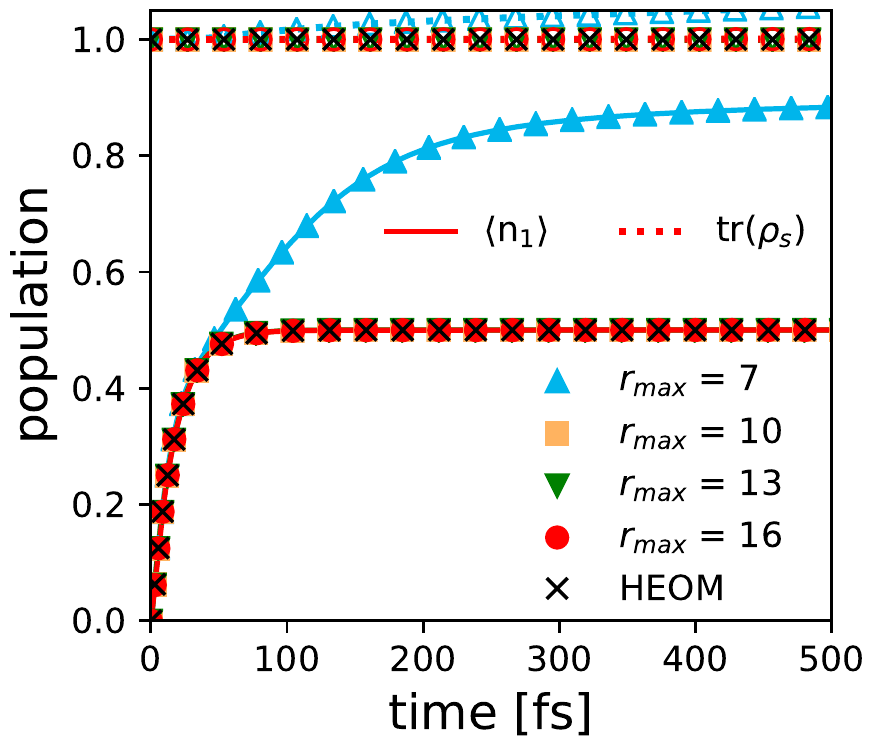}\\
			\includegraphics[width=\textwidth]{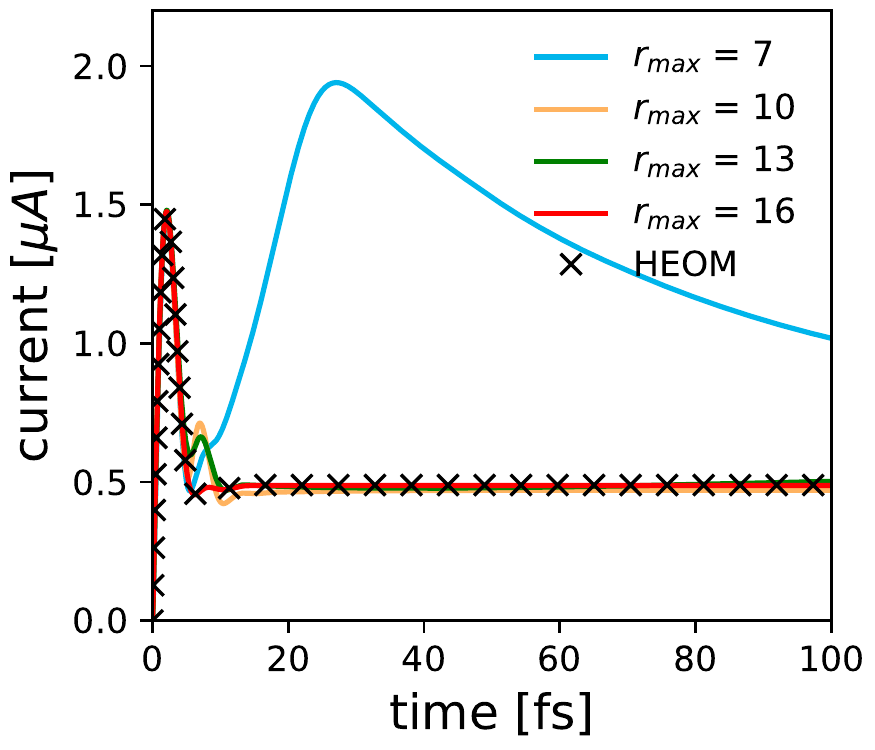}
	\end{minipage}
	\begin{minipage}[c]{0.31\textwidth}		
			\raggedright b) $\Gamma=\Delta^2=0.1$ eV\\
		\includegraphics[width=\textwidth]{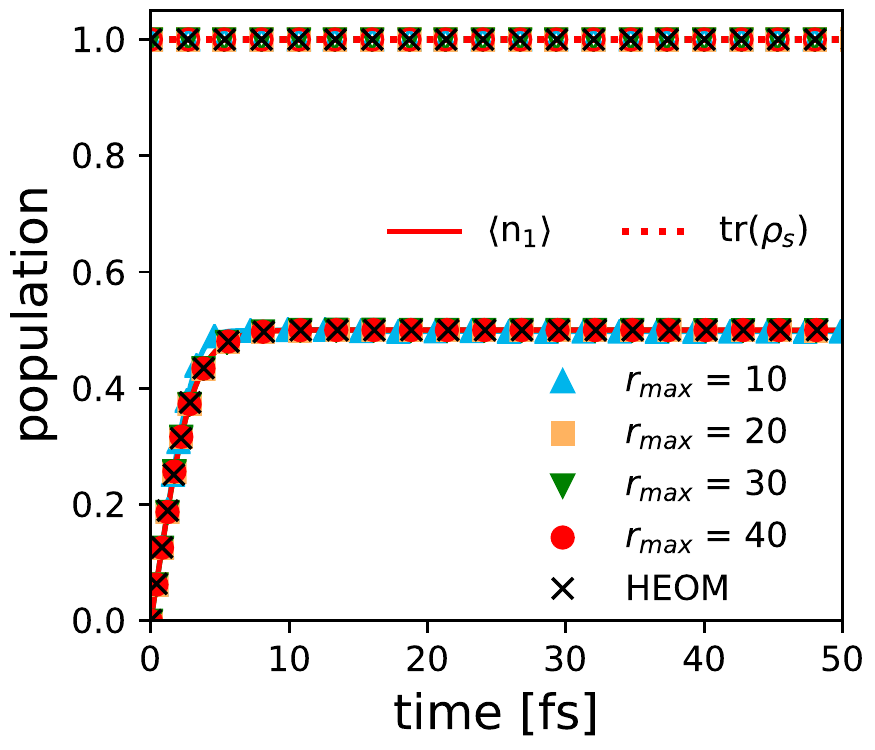}\\
		\includegraphics[width=\textwidth]{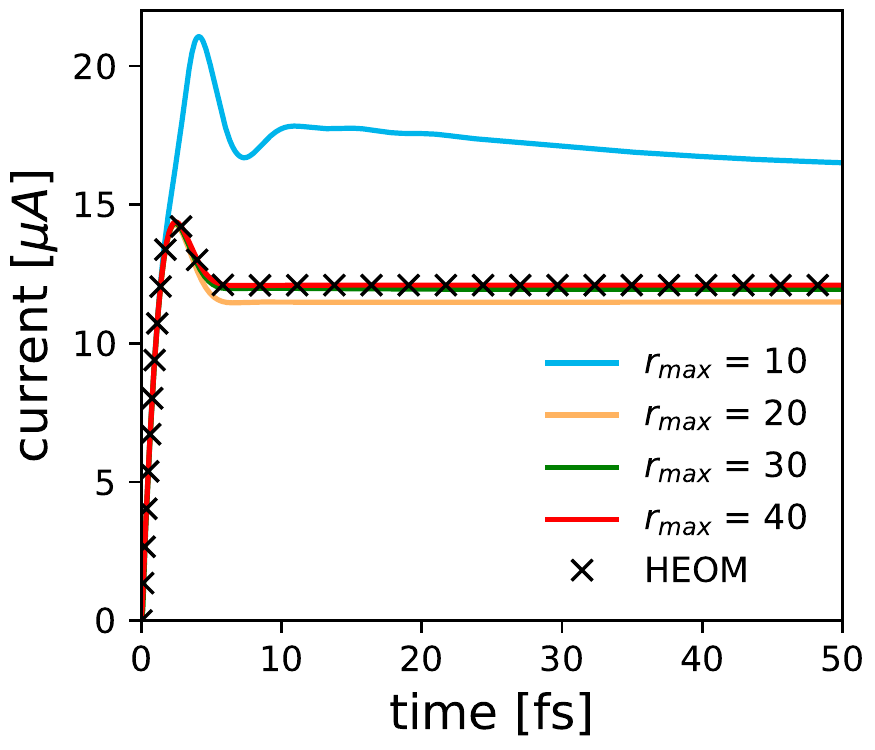}
	\end{minipage}
		\begin{minipage}[c]{0.31\textwidth}		
			\raggedright c) $\Gamma=\Delta^2=10$ eV\\
		\includegraphics[width=\textwidth]{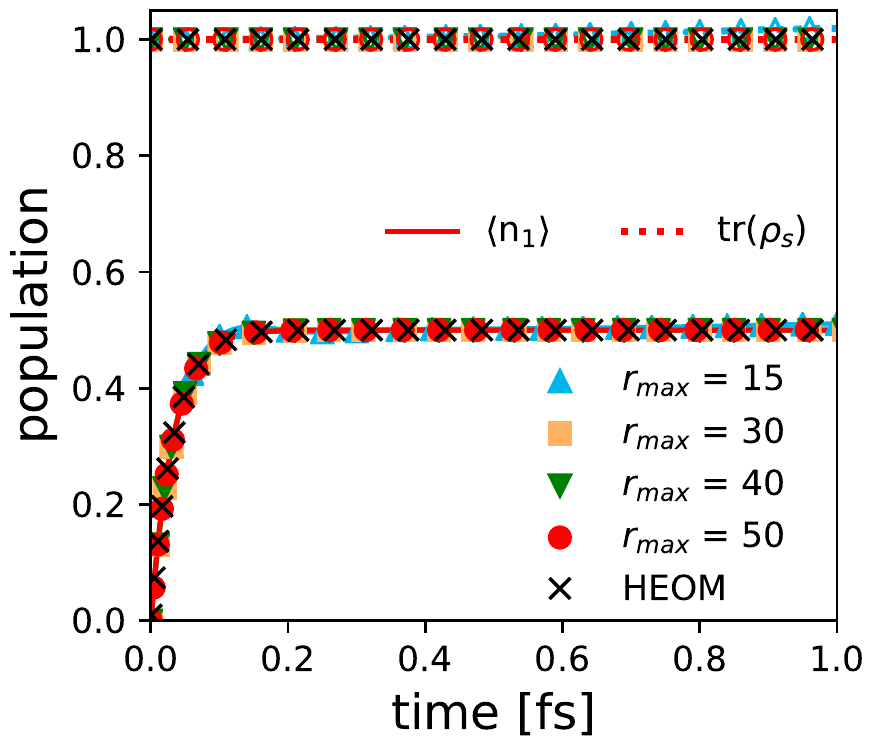}\\
		\includegraphics[width=\textwidth]{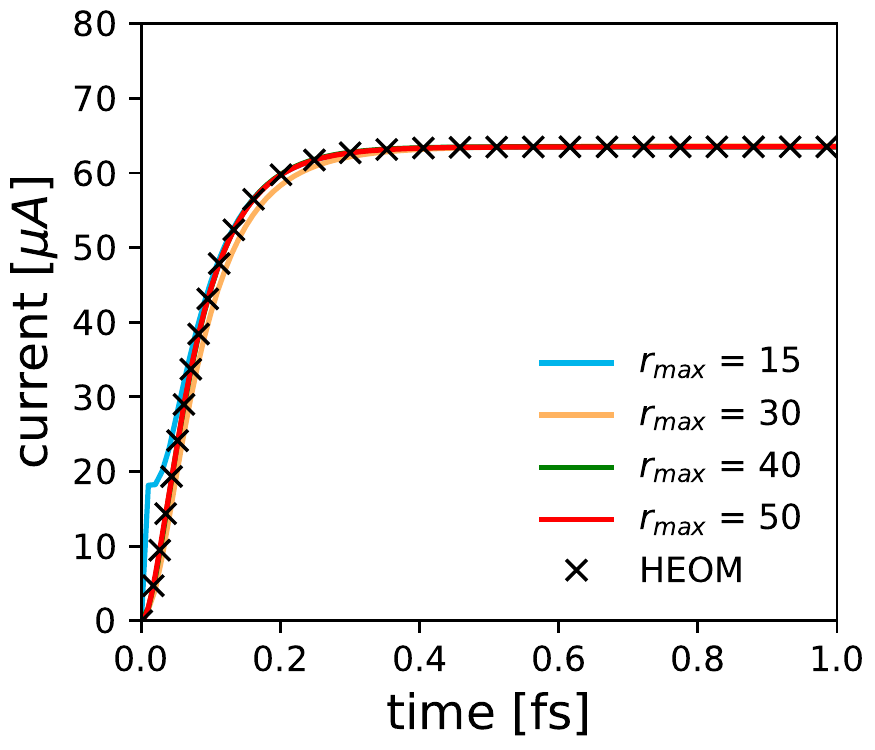}
	\end{minipage}
	\caption{Population of the first electronic level and the norm (first row), as well as the current (second row) as a function of time for three different molecule-lead couplings $\Gamma=\Delta^2=0.01$ eV (a), $\Gamma=0.1$ eV (b), and $\Gamma=10$ eV (c) in a simple electronic two-level model (see \Eq{twolevel}). The parameters of the model are :  $\epsilon_{1}=\epsilon_{2}=-0.5$ eV, $U = 1$ eV, $V=0$ eV. The temperature is $T=0.1$ eV and the bias voltage $\Phi=\mu_L-\mu_R=0.5$ V.  We invoke the wide-band approximation, and the number of fermionic Pad\'e poles is $P_{\rm{f}}=6$. The results are obtained with the time step size of 0.1 fs. We show the convergence behavior of the methodology for different maximal ranks of the cores, and as a reference, the numerically exact results obtained through the conventional HEOM method are shown as cross symbols.  }
	\label{fig2}	
\end{figure*}

The reduced system observables, such as the population of the first electronic level $\langle n_{1}\rangle$ and the norm of the reduced system density matrix $\mathrm{tr}(\rho_s)$, are obtained as the expectation value (or inner product) of
\begin{equation}
    \langle n_{1}\rangle = \langle \mathbbm{I}_0| \hat{d}_{1}^+\hat{d}_{1}^- |\Psi(t)\rangle,
\end{equation}
\begin{equation}
    \mathrm{tr}(\rho_s) = \langle \mathbbm{I}_0 |\Psi(t)\rangle,
\end{equation}
with $|\mathbbm{I}_0\rangle=| \bm{n}=\bm{0} \rangle \otimes |\mathbbm{1} \rrangle \otimes |\bm{m}=\bm{0}\rangle $ and the unit vector $|\mathbbm{1}\rrangle$ defined in \Eq{unit_vector}.
The current, which is a bath-related observable, is calculated using the formula
\begin{equation}
   I(t) =\sum_{k=1}^{K}  (\delta_{\alpha_k,L} - \delta_{\alpha_k,R})\Delta_{i_k \alpha_k} \langle \mathbbm{I}_{\bm{1}_k}|\hat{d}_{i_k}^{\sigma_k}|\Psi(t)\rangle ,
\end{equation}
with $|\mathbbm{I}_{\bm{1}_k}\rangle=| \bm{n}=\bm{1}_k \rangle \otimes |\mathbbm{1} \rrangle \otimes |\bm{m}=\bm{0}\rangle $. 

The extended state $|\Psi\rangle$ is initialized in the ground state, i.e. all the electronic levels are unpopulated:  
\begin{equation}
    |\Psi(t=0)\rangle = |\underbrace{0 \cdots 0}_{M}\rangle.
\end{equation}
Within the MPS/TT representation, and for each tensor $A^{[i]}$, we have the element $A^{[i]}(0,0,0)=1$ and all other values are set to zero. 

Although the TDVP integrator is known to be a symplectic algorithm that preserves the norm and energy during the time propagation,\cite{Yang_Phys.Rev.B_2020_p094315} this is not necessarily the case in the HEOM+TT approach. On the one hand, it is due to the fact that the super-Hamiltonian $\mathbbm{H}$ in \Eq{final_equation} is non-Hermitian. On the other, the trace of the reduced density matrix $\mathrm{tr}(\rho_s)$ is not the same as the norm of the extended wave function $|\Psi(t)\rangle$. The deviation of $\mathrm{tr}(\rho_s)$ from  unity can therefore serve as a measure for error analysis. 

\Fig{fig2} displays the time-dependent population of the first electronic level, norm, and current using the HEOM+TT approach for different maximal value of ranks.  All the related parameters are given in the caption. As a reference, the converged results obtained through the conventional HEOM approach are shown as black cross marks.

As shown in \Fig{fig2} (a) for weak molecule-lead coupling, $\Gamma=0.01$ eV, the norm $\mathrm{tr}(\rho_s)$ is well-preserved with a small maximal rank $r_{\rm{max}}=10$, and the population of the first electronic state $\langle n_1\rangle$ is in an excellent agreement with the numerically exact result obtained with the conventional HEOM approach using a fourth-tier hierarchical truncation. For the current, the converged result is obtained when the maximal rank is increased to $r_{\rm{max}}=16$. As expected, for a stronger molecule-lead coupling (which also means a stronger entanglement between molecule and leads), a larger rank is required. In this case, accurate population dynamics are obtained with $r_{\rm{max}}=20$ for $\Gamma=0.1$ eV and $r_{\rm{max}}=30$ for $\Gamma=10$ eV. The current imposes a more stringent demand on the rank and converged results are obtained with the maximal rank of tensor train cores $r_{\rm{max}}=40$ for $\Gamma=0.1$ eV and $\Gamma=10$ eV. 

We should emphasize here that the HEOM+TT method automatically contains all tiers of the fermionic hierarchy, which is especially important in cases where co-tunneling and higher-order processes play a significant role, such as for strong molecule-lead coupling or low applied bias voltage. 

In \Fig{fig2}, we present results only for a fixed bias voltage and temperature. Results corresponding to other parameters are provided in the supplementary material. It turns out that the maximally required rank is rather insensitive to the value of bias voltage and temperature. Interestingly, we also found that the maximally allowed time step for parameter sets in \Fig{fig2} can be orders of magnitude larger in the HEOM+TT approach than in the conventional HEOM approach. 

\begin{figure*}
	\begin{minipage}[c]{0.31\textwidth}		
			\raggedright a) $\Phi=0$ V \\
			\includegraphics[width=\textwidth]{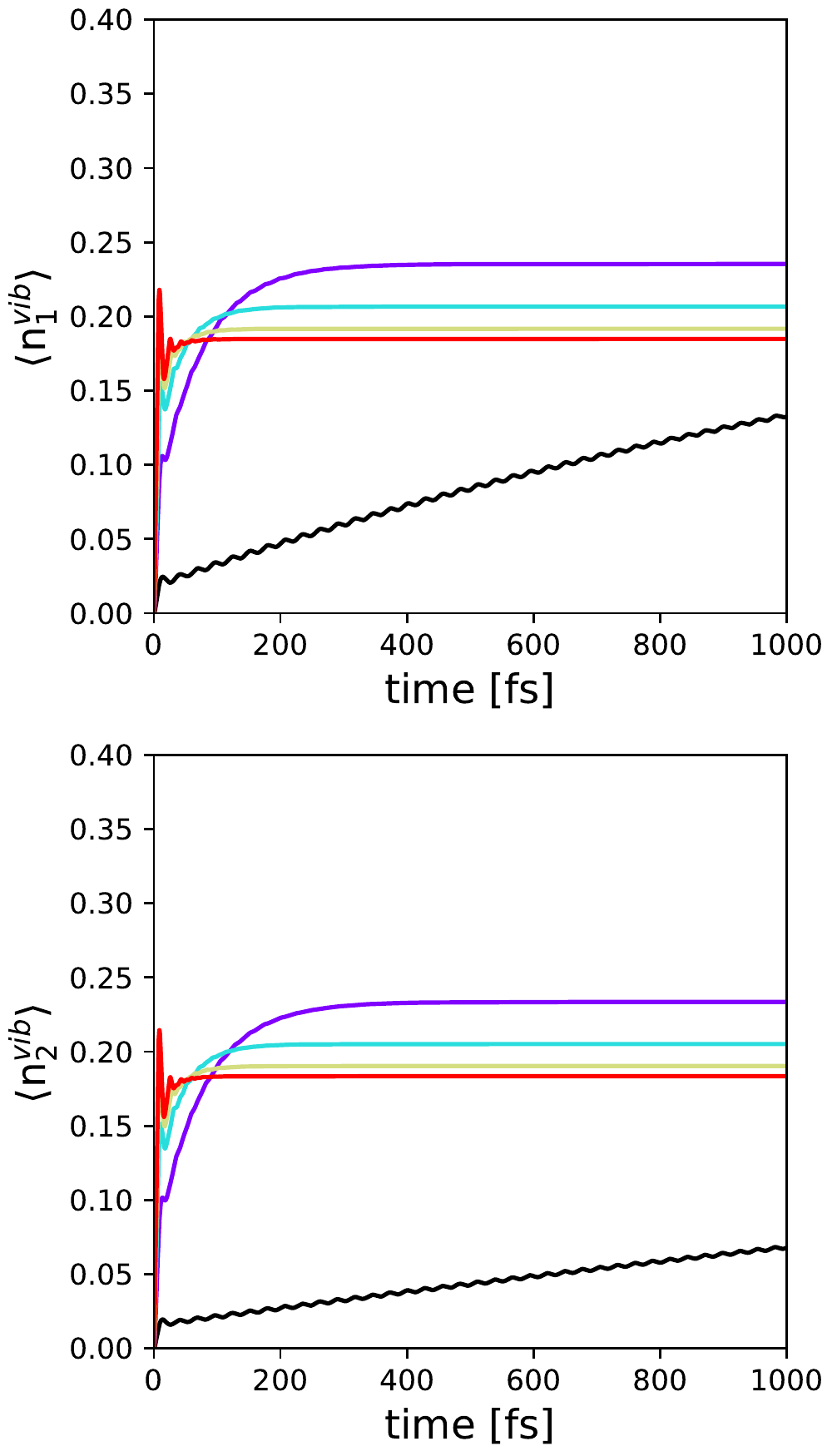}
	\end{minipage}
	\begin{minipage}[c]{0.31\textwidth}	
			\raggedright b) $\Phi=1$ V \\
			\includegraphics[width=\textwidth]{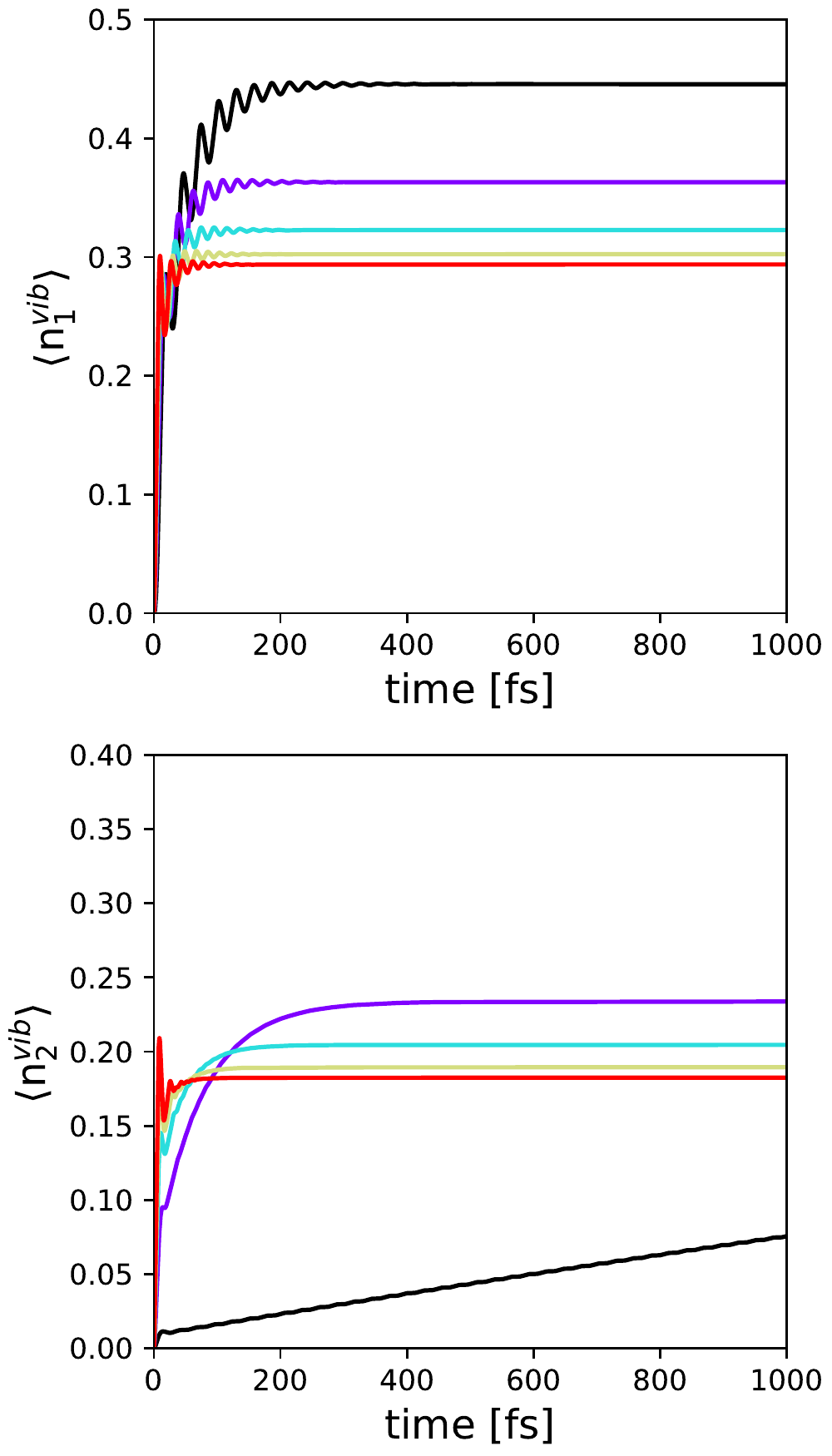}
	\end{minipage}
	\begin{minipage}[c]{0.31\textwidth}	
			\raggedright c) $\Phi=4$ V \\
			\includegraphics[width=\textwidth]{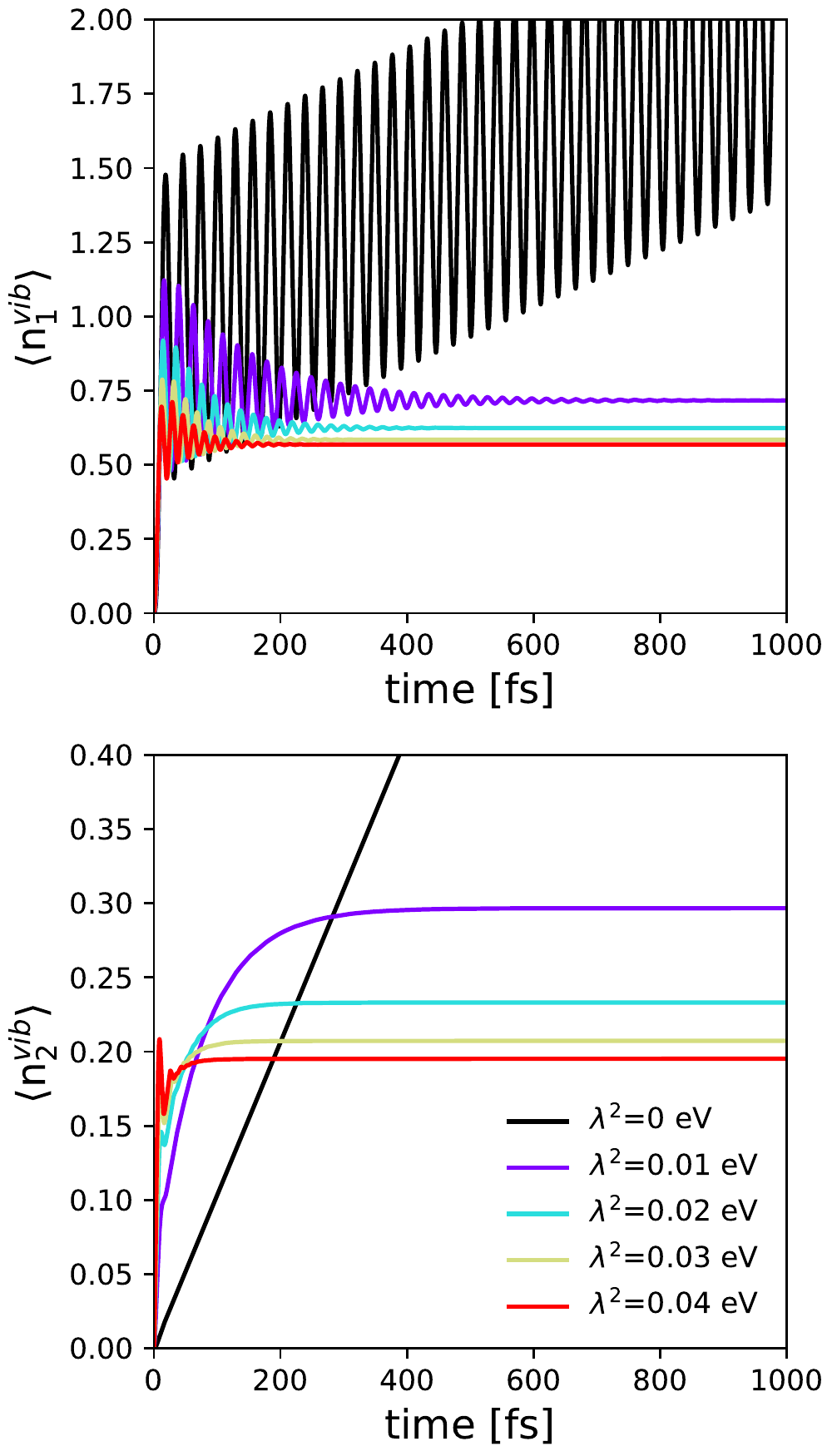}
	\end{minipage}
	\caption{Average vibrational excitation for two vibrational modes (the first shown in the left column and the second in the right column) in a molecular junction model under three different bias voltages $\Phi$.   In each panel, lines in different colors correspond to different coupling strength $\lambda$ between the vibrational modes and the phonon baths.  }
\label{fig3}	
\end{figure*}
\begin{figure}
	\begin{minipage}[c]{0.45\textwidth}		
						\raggedright a)\\			\includegraphics[width=\textwidth]{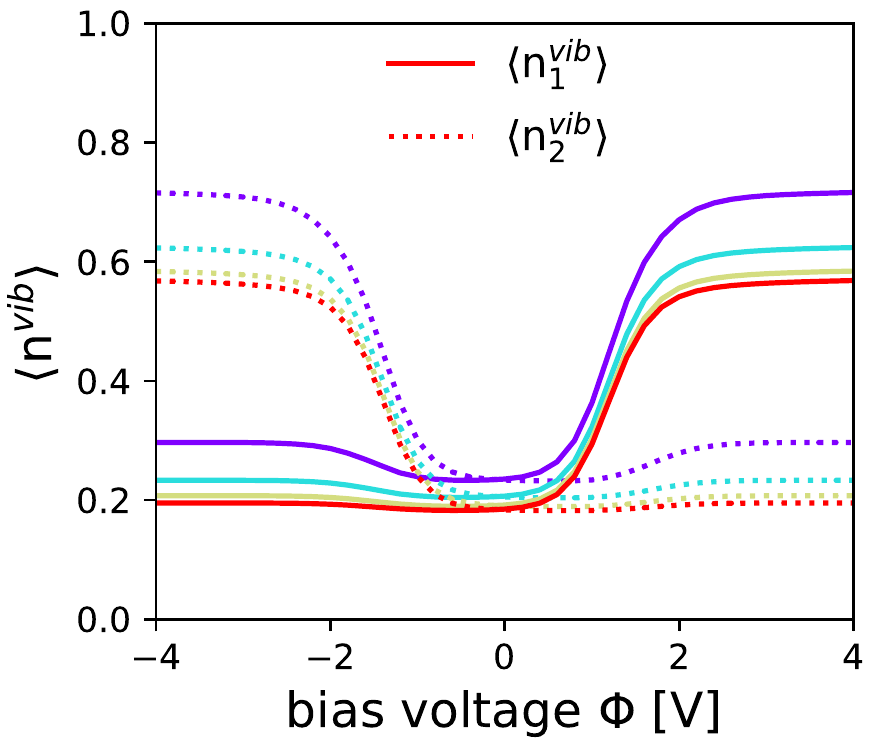}\\
						\raggedright b) \\
			\includegraphics[width=\textwidth]{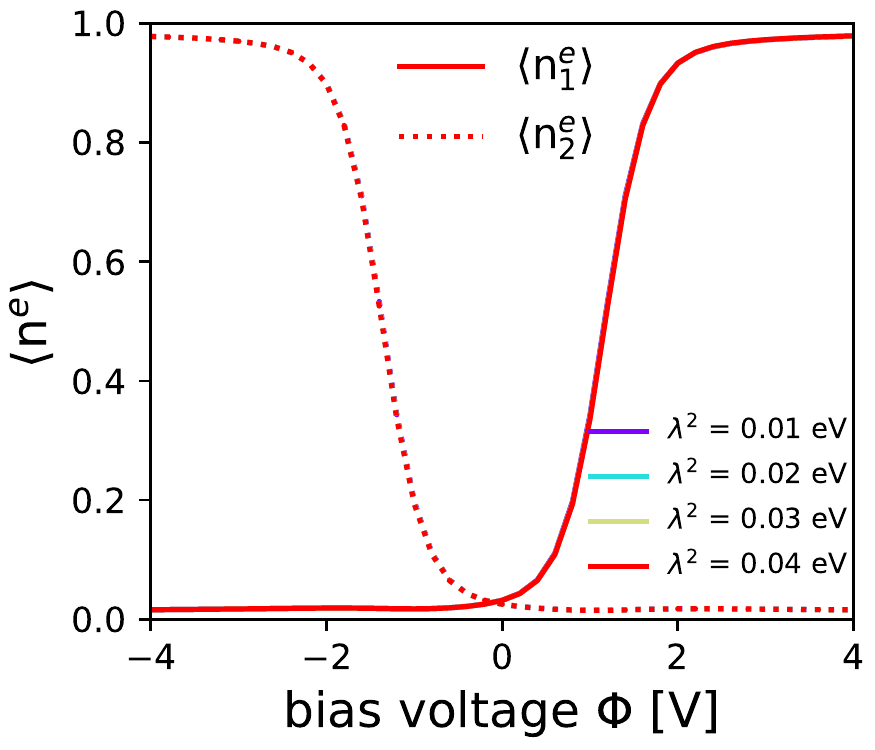}
	\end{minipage}
	\caption{Voltage-dependent mode-selective vibrational excitation (a) and the respective population of the eletronic levels (b) in an asymmetric molecular junction model as described in the main text. The results are obtained by propagating in time to reach the steady states and then extracting the plateau values.}
\label{fig4}	
\end{figure}
\subsection{Vibrational model of mode-selective vibrational excitation in molecular junctions}
As a second example, we consider a larger
system, which goes beyond the feasibility of the conventional HEOM method. To this end, we consider a model of a nanoscale molecular junction, where the system comprises two electronic levels and two internal vibrational modes
.\cite{Haertle_J.Chem.Phys._2010_p081102} Specifically, the molecular Hamiltonian reads
 \begin{equation}
 \label{mode_selective}
 \begin{split}
H_s=& \epsilon_{1} d_{1}^+d_{1}^- 
+\epsilon_{2} d_{2}^+d_{2}^-
+ Ud_{1}^{+}d_{1}^-d_{2}^{+}d_{2}^- 
+\frac{\omega_1}{2}(p_1^2+x_1^2) \\
& +\frac{\omega_2}{2}(p_2^2+x_2^2)  
+\frac{g_1}{\sqrt{2}}x_1d_{1}^+d_{1}^- 
+\frac{g_2}{\sqrt{2}}x_2 d_{2}^+d_{2}^-.
\end{split}
 \end{equation}
Here, $x_{1/2}$ and $p_{1/2}$ are the position and momentum operators for the vibrational modes with frequencies $\omega_{1/2}$. 
Each electronic level is coupled exclusively to one vibrational mode with the strength $g_{1/2}$. Every vibrational mode is coupled to its own phonon bath with coupling strength $\lambda_{1}=\lambda_2=\lambda$ and cutoff frequency $\Omega_1=\Omega_2=\Omega$ (see \Eq{lorentzian}).
Besides, every electronic level is coupled asymmetrically to two leads, one with $\Gamma_{1L}=100 \Gamma_{1R}=\Gamma$ and the other with $\Gamma_{2R}=100 \Gamma_{2L}=\Gamma$. The system can be interpreted as two electronic states being localized at different parts of the molecule. This localization can be 
tuned by attaching electron-withdrawing or -donating functional side groups at different parts of the molecule, as suggested in Ref. \onlinecite{Haertle_J.Chem.Phys._2010_p081102}. The model was proposed as a prototype for achieving mode-selective vibrational excitation in non-equilibrium scenarios by applying a finite bias voltage, $\Phi$.\cite{Haertle_J.Chem.Phys._2010_p081102} The applied bias voltage drops symmetrically on both leads, $\mu_L=-\mu_R=\Phi/2$. 
Here, we choose the cutoff frequency of the phonon baths $\Omega$ to be comparable with $\Gamma$ and the vibrational frequency $\omega_{1/2}$, which means that the correlation time of the environment is on the same scale as system electronic and vibrational dynamics. In this so-called intermediate regime,  pertubative methods are known to fail in capturing the system dynamics,\cite{Ishizaki_J.Chem.Phys._2009_p234110} and advanced numerical methods are necessary.

\Fig{fig3} shows the average vibrational excitation $ \langle n_{1/2}^{\rm{vib}}\rangle$ of the two vibrational modes,
\begin{equation}
    \langle n_{1/2}^{\rm{vib}}\rangle = \mathrm{tr}\left(\rho_s(t) \frac{\left(p_{1/2}^2 + x_{1/2}^2\right)}{2}\right) = \langle \mathbbm{I}_0| \frac{\hat{p}_{1/2}^2 + \hat{x}_{1/2}^2}{2} |\Psi(t)\rangle,
\end{equation}
as a function of time for different bias voltages $\Phi$ and coupling strength to the phonon bath $\lambda$. The system is initialized in the vibrational ground state. 
In our simulations, the wide-band approximation is invoked. We set the molecule-lead coupling strength as $\Gamma=0.1$ eV and the electron-vibrational coupling strength as $g_{1/2}=0.15$ eV. 
The two modes have the identical vibrational frequency of $\omega_{1/2}=0.15$ eV.  
Other parameters are $\epsilon_1=0.65$ eV, $\epsilon_2=0.75$ eV, $U=0$ eV, $T=0.1$ eV,  $\Omega=0.1$ eV.

Convergence of the results with respect to the number of Pad\'e poles, the size of vibrational basis set, and the maximal rank, is achieved with $P_{\rm{f}}=10, P_{\rm{b}}=5$,  $N_{\rm{vib} }=N_h=25$, and $r_{\rm{max} }=100$. In this case, the rank of the augmented wave function is $M=100$ and the number of elements to be stored is reduced from a prohibitively large value of the order of $10^{48}$ for the full tensor down to a manageable one of roughly $10^{5}$ in the MPS/TT format.

At zero bias voltage, as shown in \Fig{fig3} (a), the average vibrational excitation corresponds mostly to the thermal excitation induced by coupling to the electrodes and the phonon baths.  The average vibrational excitation is lower for a stronger coupling to the phonon baths, as is known for the simpler problem of a damped harmonic oscillator.\cite{grabert1988quantum}  
The excitation exhibits little difference between the two modes, because the vibrational frequencies are the same, $\omega_1=\omega_2$. 

In the non-resonant transport regime, as illustrated in \Fig{fig3} (b) for  bias voltage $\Phi=1$ V, the mode-selectivity of the vibrational excitations is evident. While for the second vibrational mode no significant difference of $\langle n_2^{\rm{vib} }\rangle$ at $\Phi=0$ and $\Phi=1$ V can be seen, $\langle n_1^{\rm{vib}}\rangle$ is significantly enhanced at $\Phi=1$ V. 

In the deep resonant transport regime, e.g. at $\Phi=4$ V (see \Fig{fig3} (c)), $\langle n_1^{\rm{vib}}\rangle$ increases further. In particular, for vanishing coupling $\lambda$, the vibrational dynamics (black line) showcases a strong and long-lasting oscillating behavior, which is an indication of coherent vibrational motion as the period of oscillations corresponds to the vibrational frequency, $2\pi/\omega_{1/2}=27.6$ fs. Moreover, the average vibrational excitation $\langle n_1^{\rm{vib}}\rangle$ is very high (data not shown). The reason behind this is the lack of dissipation because the electron-hole pair creation processes are blocked at high bias voltages.\cite{Haertle_2013_PhysicaStatusSolidib_p2365} Once dissipation to the phonon baths is introduced, the oscillations decay faster due to dissipation and $\langle n_1^{\rm{vib}} \rangle$ saturates at a finite value. Similarly, $\langle n_2^{\rm{vib}}\rangle$ continues to grow over time when $\lambda=0$, due to the absence of dissipation. When the dissipation is relatively weak ($\lambda^2=0.01$  eV), a slight increase of $\langle n_2^{\rm{vib}}\rangle$ is observed with increasing bias voltage. But for large $\lambda^2=0.04$ eV, $\langle n_2^{\rm{vib}}\rangle$ remains basically invariant for different bias voltages, implying the suppression of phonon excitations due to the strong dissipation effect. 

To elucidate the bias-controlled mode-selectivity, \Fig{fig4} shows the steady-state average vibrational excitation $\langle n^{\rm{vib}}_{1/2}\rangle$ as well as the populations of two electronic levels $\langle n^e_{1/2}\rangle $  against the bias voltage. An increase is observed in the bias regime $\Phi=0$ to $2$ V for both $\langle n_1^{\rm{vib}}\rangle$ and $\langle  n_1^e\rangle $, which is followed by a saturation for $\Phi>2$ V.  Reversing the bias polarity, the vibrational populations of the two modes are inverted, as are the electronic populations. Moreover, we notice that the mode-selectivity is immediately activated once the applied bias voltage is turned on. 

As discussed in Refs. \onlinecite{Haertle_J.Chem.Phys._2010_p081102} and \onlinecite{Volkovich_Phys.Chem.Chem.Phys._2011_p1433314349}, the bias-controlled mode-selectivity and the above observations can be accounted for by the asymmetry of the coupling of the two electronic levels to the leads and the internal vibrational modes. In the positive bias direction, electrons tunnel from the left lead to the molecule and subsequently to the right lead. The electrons predominantly populate the first electronic level because of its strong coupling the left lead and rather weak coupling to the right lead, which is exactly the opposite for the second electronic level that is barely occupied. Because the first electronic level is exclusively coupled to the first vibrational mode, energy is directed into the first vibrational mode and thus we observe a high vibrational excitation $\langle n_1^{\rm{vib}}\rangle$ but quite small $\langle n_2^{\rm{vib}}\rangle$. For the same reason, in the negative bias direction, the second electronic level is populated from the right lead much faster than it is depopulated to the left lead, and therefore it is nearly fully occupied as long as the bias voltage is large enough. In this case, the excitation in the second vibrational mode is dominant. 

It should be emphasized that all higher-order processes, which correspond to all higher-order fermionic hierarchical tiers and which are critical in the non-resonant regime, are incorporated in our approach. As opposed to the second-order treatment of the molecule-lead coupling which predicts an onset bias of mode-selectively in the resonant regime,\cite{Haertle_J.Chem.Phys._2010_p081102,Volkovich_Phys.Chem.Chem.Phys._2011_p1433314349} we found that the mode-selectivity is pronounced even at low bias voltages.

The above results remain true for a finite Coulomb interaction $U$ and a larger cutoff frequency $\Omega$. The relevant results are presented and analyzed in the supplementary material. However, the mechanism of bias-controlled selective vibrational excitation can be significantly more complex and intriguing in cases, where, for example, the vibrational frequencies are different, the vibronic coupling as well as the coupling to the leads are very strong, and intra-molecular vibrational energy redistribution is involved.  A systematic investigation of selective vibrational excitation induced by nonequilibrium inelastic transport processes, or the selectivity of chemical reactions in realistic systems, is an interesting topic for future work. 

We emphasize that the above model goes beyond the ability of the conventional HEOM approach.
Although the HEOM method is capable to treat discrete vibrational modes in transport scenarios,\cite{Schinabeck_2016_Phys.Rev.B_p201407,Schinabeck_2018_Phys.Rev.B_p235429} it can exhibit stability issues if the modes are coupled to bosonic reservoirs.\cite{Dunn_J.Chem.Phys._2019_p184109} In the HEOM+TT approach used here, this problem is overcome by 
increasing $N_h$ without significant increase of the numerical effort.
In the supplementary material, we demonstrate the numerical performance as well as the run time of the approach with respect to the bosonic hierarchy $N_h$.

\section{Conclusion}
In this work, we have presented a highly efficient method for modeling the dynamics of open quantum system embedded in a hybrid fermionic and bosonic environment.  The key is to reformulate the numerically exact HEOM approach for the auxiliary density matrices into a time-dependent Schr\"odinger-like equation for an augmented wave function, where the system is represented in twin space and the manifold of fermionic leads and phonon baths are mapped into a finite number of discrete virtual electronic levels and vibrational modes, respectively. To facilitate practical simulations, the augmented wave function is represented in the MPS/TT formalism and propagated using the TDVP technique. 

We first benchmarked our results for a simple electronic two-level model against the ones obtained by the conventional HEOM method, to demonstrate the accuracy of the proposed method. Furthermore, to illustrate the capability of the method for more complex systems over a broad range of parameters, we studied the mode-selective vibrational excitation in an asymmetric molecular junction.
Further applications to modelling chemical reactions in molecular junctions will be the subject of  future work.

We should point out that, although the proposed method has opened up the possibility for studying a wider variety of the problems in chemical physics and beyond, there is still space for further improving the efficiency of the method. One can apply other advanced techniques developed both in the context of HEOM and MPS/TT, respectively. For example, the difficulty of applying the HEOM approach to very low temperatures, which requires a large number of poles in the spectrum decomposition of Fermi-Dirac/Bose-Einstein distribution functions, may be overcome by the Fano decomposition scheme.\cite{Zhang_2020_J.Chem.Phys._p64107} Furthermore, we have arranged all the DoFs in a one-dimensional chain, which may not be an optimal ordering and structure since there is no direct coupling between neighboring environmental modes.  A more suitable structure of the tensor train would reflect the natural structure of the super-Hamiltonian for the system of concern, which could be achieved by arranging highly entangled DoFs as close as possible and reducing the ranks to the smallest.\cite{Yan_J.Chem.Phys._2021_p194104,Kloss_SciPostPhysics_2020_p070} In addition, since the entanglement grows as time evolves, techniques that introduce rank adaptivity into the TDVP integration have been proposed,\cite{Yang_Phys.Rev.B_2020_p094315,Borrelli_J.Phys.Chem.B_2021_p53975407} which appear promising in practical applications.

\section*{Acknowledgements}
The authors thank Samuel Rudge and Jakob B\"atge for a critical reading of the manuscript and helpful discussions. MT thanks Uri Peskin for many insightful discussions on mode-selective vibrational excitation. This work was supported by the German Research Foundation (DFG).
Y.K. was supported by the Alexander von Humboldt Foundation. 
Furthermore, the authors acknowledge support by the state of Baden-Württemberg through bwHPC
and the German Research Foundation (DFG) through grant no INST 40/575-1 FUGG (JUSTUS 2 cluster).

\section*{Supplementary Material}
See the supplementary material for the  details  of  (1) coefficients and exponents in \Eq{corr_f_exp} and \Eq{corr_b_exp}; (2) different forms of the hierarchical equations of motion in the MPS/TT format and their numerical performances; (3) convergence analysis of the HEOM+TT method with respect to the bosonic hierarhcy $N_h$; (4) comparison of \Eq{final_equation} in Liouville space and twin space;  (5) results of simple  electronic two-level model with different temperatures and applied bias voltages from those in \Fig{fig2}; (6) results of the mode-selective vibrational excitation model in \Eq{mode_selective} for different Coulomb interactions and cut-off frequencies from those in \Fig{fig3}.

\section*{Data Availability}
The data that support the findings of this study are available from the corresponding author upon reasonable
request.

\end{document}